# All Dielectric Biaxial Metamaterial: A New Route to Lossless Dyakonov Waves with Higher Angular Existence Domain

Ayed Al Sayem

Dept. of EEE, Bangladesh University of Engineering and Technology, Dhaka, Bangladesh

**Abstract.** In this article, we propose and numerically analyze an all dielectric biaxial metamaterial [ADBM] constructed by multilayer pattering of a sub-wavelength ridge array of Silicon and a flat $SiO_2$ layer. The proposed ADBM can support Dyakonov Surface Waves [DSWs] with infinite propagation length which can propagate in a wide angular domain. Though natural uniaxial and biaxial materials and also nanowire all dielectric metamaterials can also support DSWs, the angular existence domain [AED] is limited to a very narrow range. Our proposed ADBM can support can overcome this limitation and it can achieve higher AED than any all dielectric structures reported in literature till date. Our proposed ADBM can be easily fabricated by the current fabrication technology. Due to its lossless nature, it may find substantial applications in optical sensing, optical interconnects, wave-guiding, solar energy harvesting etc.

## I. INTRODUCTION

Dyakonov waves are hybridized surface waves which can propagate at the interface between an isotropic dielectric medium and a uniaxial dielectric medium [1]. Though D'yakonov theoretically predicted such surface waves more than 29 years ago such waves have been experimentally demonstrated just few years back [2, 3]. DSWs can also exist at the interface between two uniaxial dielectric mediums [4], an isotropic dielectric medium and a biaxial dielectric medium [4-7], a uniaxial dielectric medium and a biaxial dielectric medium [5] and also between two biaxial dielectric mediums [5]. Different multilayer structures and metamaterials [8-13] have also been used to demonstrate DSWs. DSWs are exciting for various applications as they can propagate at the interface between two lossless mediums allowing infinite propagation length [1]. But DSWs suffer from very narrow AED [1-7] (i.e. the range of angles at the interface DSWs are allowed to propagate). This occurs mainly due to the very small anisotropy of the available natural anisotropic uniaxial and biaxial materials found in nature [1-7]. This is consequently the main cause of difficulty faced in experimental demonstrations of DSWs along with the excitation of DSWs. With the decade long extreme investigations on metamaterials [14-16], they have been also used in DSW research. Especially hyperbolic materials either natural [17] or artificial [18, 19] can increase the AED. But such natural hyperbolic materials are lossy [17, 20] and hyperbolic metamaterials contain metals or graphene which are also lossy [21-26]. As a result, the lossless property of DSWs is completely lost. All dielectric uniaxial metamaterials [27, 28] such as 2D photonic crystals with deep sub-wavelength unit cells have been demonstrated to have a much higher AED than natural ones [8]. But the fabrication and especially the orientation of such horizontally oriented nanowire structures are quite complicated.

In this article, we theoretically propose and numerically demonstrate an ADBM which can possess much higher AED than natural ones and also uniaxial all dielectric metamaterials [8]. Such an ADBM can be constructed by a multilayer structure where each unit cell of the structure contains a deep sub-wavelength dielectric ridge array and a planer dielectric layer as shown in Fig. 1. Using the parameter retrieval procedure [29-31], the permittivity tensor components of the proposed ADBM have been obtained and DSWs and especially the AED achievable by the proposed structure have been analyzed. The proposed ADBM which can support DSWs with infinite propagation length and much wider AED may find substantial applications in optical sensing [32], wave guiding [33], interconnects, solar energy harvesting [34] etc.

## II. THEORY & NUMERICAL SIMULATIONS

The schematic of the proposed structure is shown in Fig. 1. The unit cell of the structure contains a sub-wavelength Silicon ridge (infinite along z’ axis) array with period w along the y’ axis and a flat $SiO_2$ layer. The thickness and width of the silicon ridge is $t_1$ and $w_1$ respectively. The thickness of the flat silicon layer is $t_2$. The unit cell is repeated along the x’ direction with period $t=t_1+t_2$. Above the structure, an isotropic dielectric medium with permittivity, $\varepsilon_d$ is placed at the upper half-space x’> 0. Here, we note that the first layer (at the interface of isotropic clad medium and ADBM) of the ADBM can be either the ridge array or the flat $SiO_2$ layer. This doesn’t alter the results of this article. The interface between the isotropic dielectric medium and ADBM is the y’-z’ plane. DSWs can propagate along the z axis which makes an angle θ with the z’ axis.

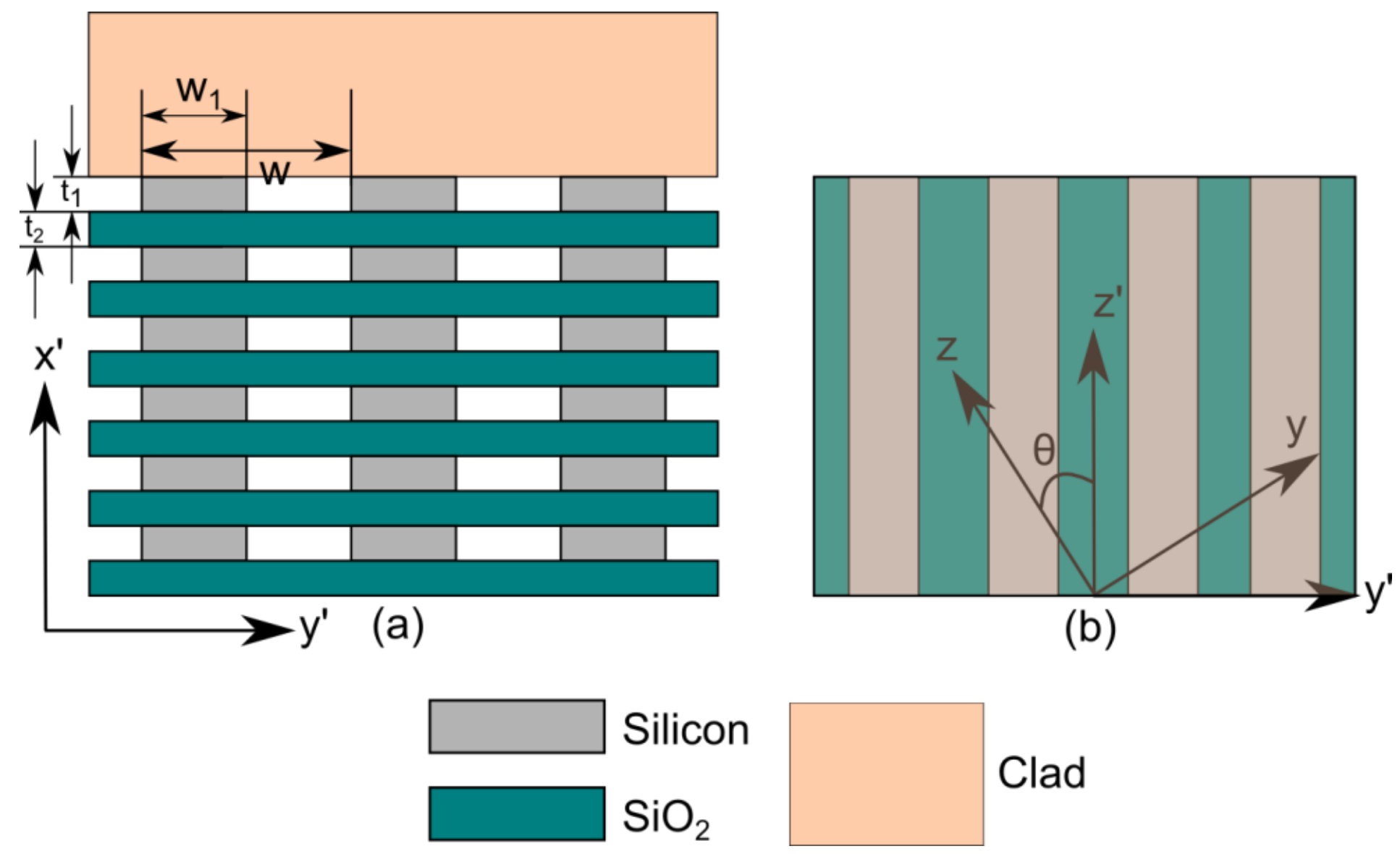

Fig. 1 Schematic of the proposed ADBM (a) side view (x'-y' plane) (b) top view (z'-y' plane). Silicon ridge with width, $w_1$ and thickness, $t_1$ is repeated along the y' axis with period w on the top of a flat $SiO_2$ layer with thickness $t_2$. The array of silicon ridge and flat $SiO_2$ layer is repeated along the x' axis with period $t=t_1+t_2$. The void regions contain air with permittivity, $\varepsilon_{air} = 1$. Above the ADBM, the upper half space is filled with an isotropic clad medium with permittivity, $\varepsilon_d$. DSWs propagate along the z axis which makes an angle θ with the z' axis. The parameter values except $w_1$ which have been used in all the numerical calculations of this article, w=20nm, $t_1=t_2=10$nm, permittivity of Silicon, $\varepsilon_{Si}=12.09$ and permittivity of $SiO_2$, $\varepsilon_{SiO2}=2.085$ and operating wavelength, λ=1.55μm.

For a biaxial metamaterial, in the (x', y', z') co-ordinate system, the permittivity tensor has the following form,

$$\boldsymbol{\varepsilon} = \begin{bmatrix} \varepsilon_x & 0 & 0 \\ 0 & \varepsilon_y & 0 \\ 0 & 0 & \varepsilon_z \end{bmatrix} \quad (1)$$

Where, $\varepsilon_x \neq \varepsilon_y \neq \varepsilon_z$

Using the general approach of parameter retrieval procedure [29-31], we have obtained the effective permittivity and permeability tensor values of the proposed ADBM. Fig. 2(a) shows the effective permittivity tensor values of the proposed ADBM as function of ridge width, $w_1$. We have found that permeability tensor values are close to unity (cf. supplementary information for details). Hence, we have treated the structure as non-magnetic. Using Comsol Multiphysics Finite Element Method (FEM) simulation, at first we have obtained the complex transmission and reflection coefficients of the proposed ADBM for both TE (s) and TM (p) polarization. Using the complex reflection and transmission coefficients, the permittivity and permeability tensor values for the proposed structure have been retrieved [29-31] (cf. supplementary information for details of the parameter retrieval procedure). Fig. 2(b) shows the transmission and reflection coefficients of the proposed structure at λ=1.55µm for both TE (s) and TM (p) polarization as a function of ridge width, $w_1$.

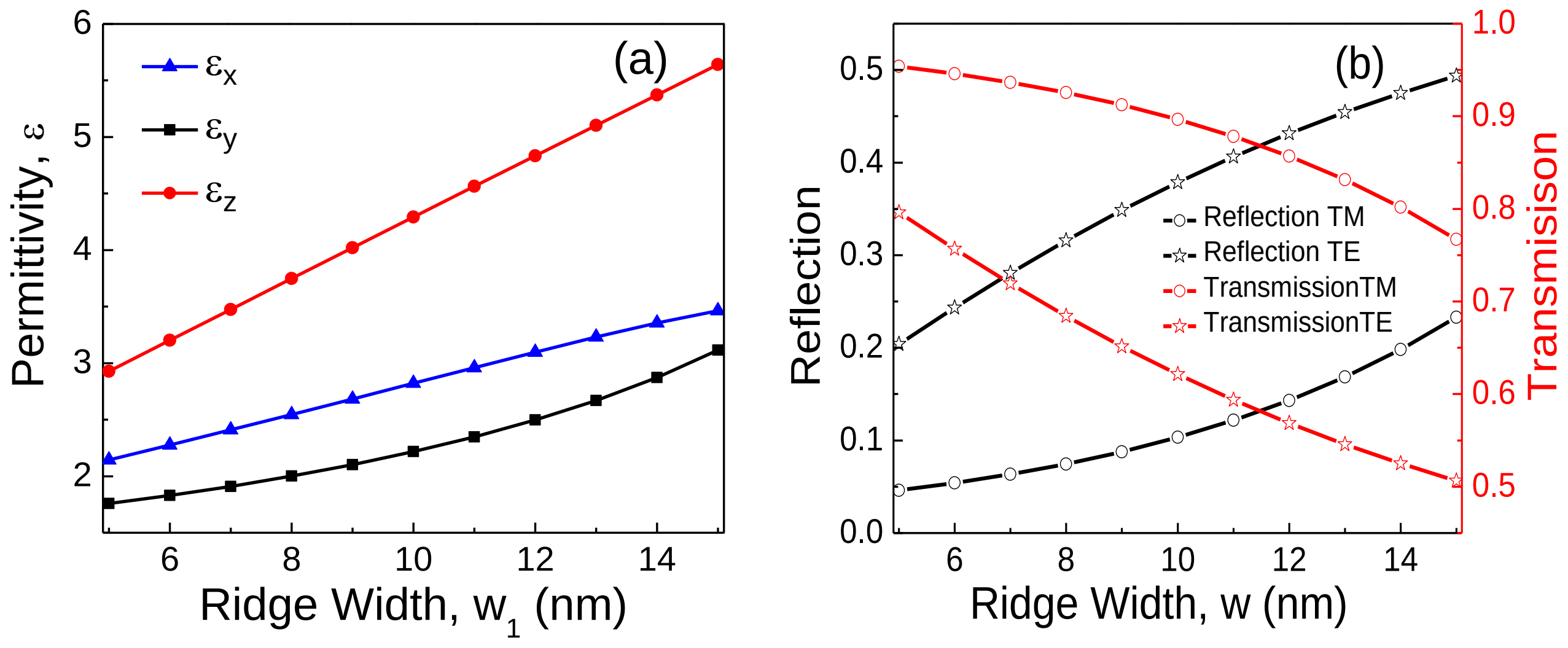

Fig. 2(a) Retrieved optical parameters (effective permittivity tensor values) of the proposed structure as a function of ridge width, $w_1$ (b) Transmission and reflection coefficients of the proposed ADBM as a function of ridge width, $w_1$ at λ=1.55µm for both TE (s) and TM (p) polarization. Total thickness of the structure is 160nm. All other parameters are same as mentioned in Fig. 1.

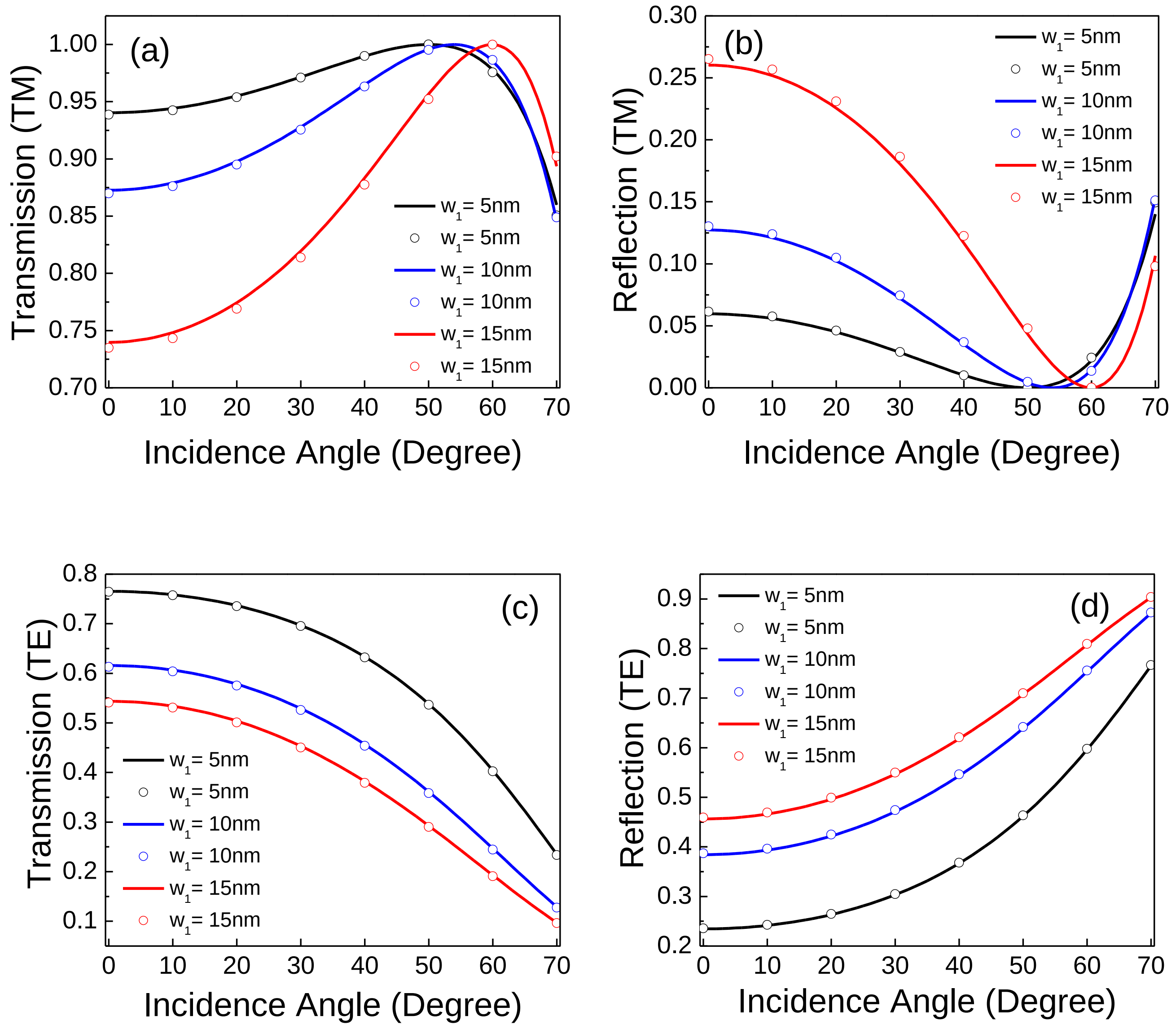

Fig. 3 Transmission and reflection coefficients for the (a, b) TM (p) and (c, d) TE (s) polarization as a function of incidence angle at λ=1.55μm for different values of ridge width, $w_1$. Solid lines and symbols represent the values calculated by theoretical calculations using the retrieved permittivity tensor values and Comsol Multi-physics simulations respectively. Total thickness of the structure is 200nm. All other parameters are same as mentioned in Fig. 1.

Figs. 3(a) and 3(b) show the transmission and reflection coefficients for TM (p) polarization as a function of incidence angle at λ=1.55μm for different values of ridge with, $w_1$. Solid lines and symbols represent the values calculated by theoretical calculations using the retrieved permittivity values (as shown in Fig. 1) and Comsol Multi-physics simulations respectively. Figs. 3(c) and 3(d) show the same but for TE (s) polarization. Excellent agreement between theoretical calculations and full wave simulations can be observed in Figs. 3(a)-3(d) which completely validates that the proposed structure behaves as a biaxial metamaterial.

The dispersion relation for the DSW at the interface of an isotropic and biaxial medium can be given by [7],

$$k_1 k_2 F - k_d^2 D - \varepsilon_d C = k_d (k_1 + k_2)(D + \varepsilon_d k_1 k_2) \tag{2}$$

Where,

$$k_j^2 = \frac{B \pm (B^2 - 4C)^{\frac{1}{2}}}{2}; j = 1, 2$$

$$B = q^2 \left(\frac{\varepsilon_{zz}}{\varepsilon_x} + 1\right) - \varepsilon_y - \varepsilon_z$$

$$C = D\left(\frac{q^2}{\varepsilon_x} - 1\right)$$

$$D = q^2 \varepsilon_{zz} - \varepsilon_y \varepsilon_z$$

$$F = \varepsilon_d(\varepsilon_y + \varepsilon_z - q^2) - \varepsilon_{zz} q^2$$

$$k_d^2 = q^2 - \varepsilon_d$$

$$\varepsilon_{zz} = \varepsilon_y \sin^2\theta + \varepsilon_z \cos^2\theta$$

Here, we have assumed that a monochromatic DSW with wave vector (0, 0, q) is propagating along the z axis. All the wave vectors have been normalized by $k_0 = \frac{2\pi}{\lambda}$ where $\lambda$ is the operating wavelength. The DSW is composed of superposition of the TE (s) and TM (p) polarized waves with same wave vector component, q in the z direction [1-7].

The lower and higher cut-off angles for which DSWs are allowed to propagate can be given by [7],

$$\theta_{min} = \sin^{-1}\sqrt{\frac{\varepsilon_d - \varepsilon_x}{\varepsilon_z - \varepsilon_y}\left(\frac{\varepsilon_d - \varepsilon_x}{\varepsilon_z - \varepsilon_y} - \frac{\varepsilon_d}{2\varepsilon_x} + \left(\frac{\varepsilon_d^2}{4\varepsilon_x^2} + \frac{(\varepsilon_d - \varepsilon_y)(\varepsilon_z - \varepsilon_d)}{\varepsilon_x(\varepsilon_d - \varepsilon_x)}\right)^{\frac{1}{2}}\right)} \tag{3a}$$

$$\theta_{max} = \sin^{-1}\sqrt{\frac{\varepsilon_z^2(\varepsilon_d - \varepsilon_y)G}{(\varepsilon_z - \varepsilon_y)\left(\varepsilon_d\varepsilon_z G - \varepsilon_y(\varepsilon_d - \varepsilon_y)(\varepsilon_z - \varepsilon_d)^2\right)}} \tag{3b}$$

Where, $G = \varepsilon_d(\varepsilon_z - \varepsilon_d) + \varepsilon_y\left(\frac{\varepsilon_d^2}{\varepsilon_x - \varepsilon_z}\right)$

The AED for which DSWs can propagate can be given by,

$$\Delta\theta = \theta_{max} - \theta_{min} \tag{4}$$

Equations (3-4) can be also used for calculating AED for uniaxial nanowire metamaterials, uniaxial hyperbolic metamaterials [7, 8, 18] using $\varepsilon_z = \varepsilon_{||}$ and $\varepsilon_x = \varepsilon_y = \varepsilon_{\perp}$ where $\varepsilon_{||}$ and $\varepsilon_{\perp}$ are the permittivity values along the parallel and perpendicular direction of the nanowires respectively.

For the co-ordinate system shown in Fig. 1, DSWs can exist only for three cases [7], $i)\ \varepsilon_z > \varepsilon_d > \varepsilon_y \geq \varepsilon_x$ $ii)\ \varepsilon_x > \varepsilon_d > \varepsilon_y \geq \varepsilon_z$ $iii)\ \varepsilon_z > \varepsilon_d > \varepsilon_x \geq \varepsilon_y$. From the permittivity tensor values shown in Fig. 2(a), one can observe that case iii) can be satisfied for our proposed structure. Also it has already been found in Ref. [7] that for case iii) highest AED can be achieved. Here, we define a dimensionless parameter $\xi$ for ADBM as,

$$\xi = \frac{\varepsilon_d - \varepsilon_x}{\varepsilon_z - \varepsilon_x} \tag{5}$$

The parameter $\xi$ can have any value between 0 and 1.

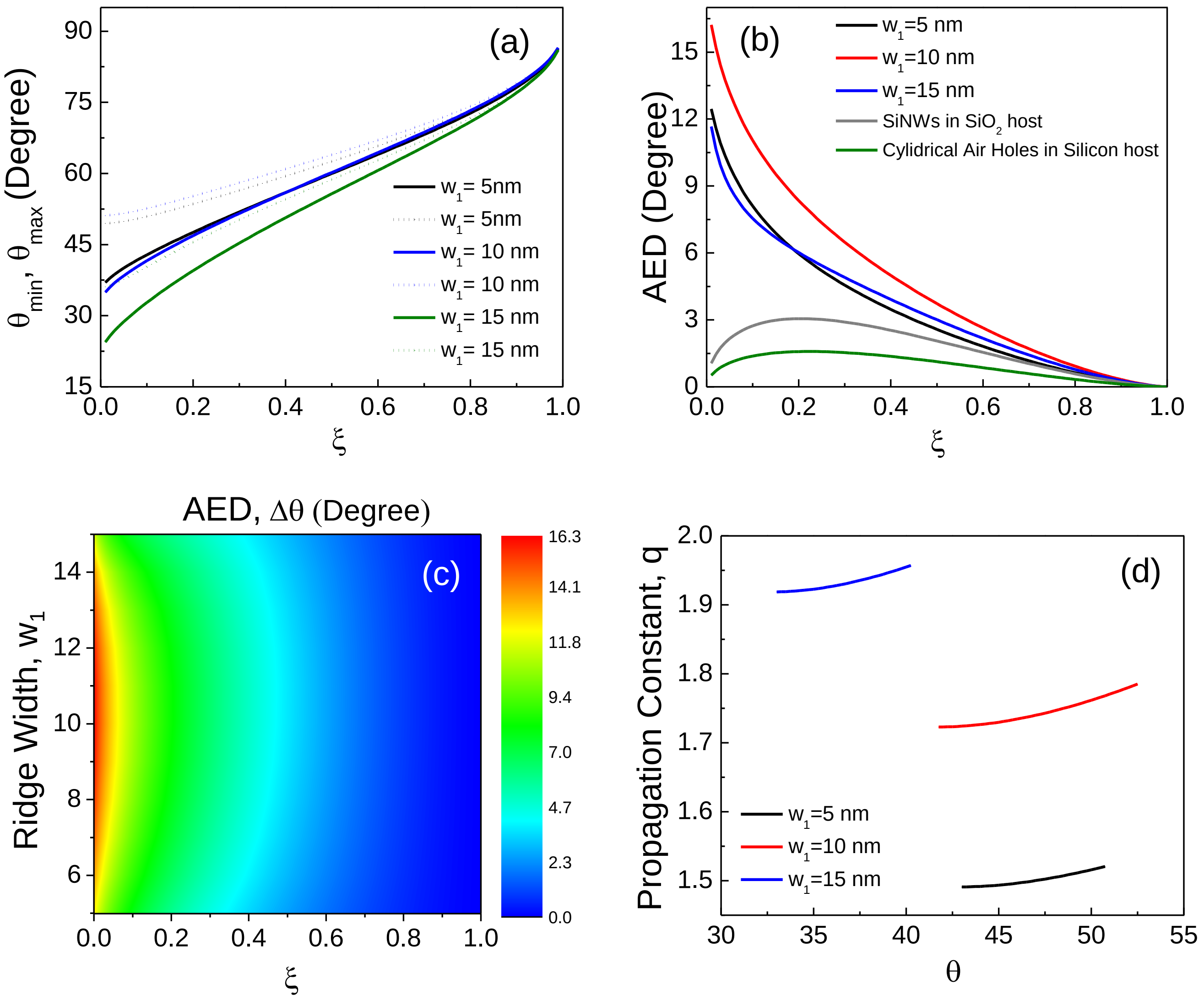

Fig. 4(a) Limits of the AED, θ (solid lines and dotted lines represent $\theta_{min}$ and $\theta_{max}$ respectively) between which DSWs are allowed to propagate as a function of $\xi$ for different values of ridge width, $w_1$ (b) AED as a function of $\xi$ for different values of ridge width, $w_1$ for the ADBM and SiNWs in $SiO_2$ host and cylindrical air holes in Silicon host (c) 2D map of the AED as a function of ridge width, $w_1$ and $\xi$ (d) Propagation constant, q as a function of rotation angle, θ for different values of ridge width, $w_1$. All other parameters are same as mentioned in Fig. 1.

Fig. 4(a) shows the limits of the AED, $\theta_{min}$ and $\theta_{max}$ as a function of $\xi$ for different values of ridge width. Fig. 4(b) shows the AED as function of $\xi$ for different values of ridge width. Along with ADBM, AED for Silicon Nanowires [SiNWs] in $SiO_2$ host and cylindrical air holes in Silicon host are also shown in Fig. 4(b) for comparison. Such SiNWs in $SiO_2$ host or cylindrical air holes is Silicon host can behave as a uniaxial dielectric medium in the long wavelength limit [8, 35] and have been previously used for increasing the AED of DSWs [8]. For SiNWs and air holes, the value of fill factor was fixed at 0.43 and 0.73 respectively (cf. supplementary information for details). It can be observed from Fig. 4(b) that for a wide range of values of $\xi$ (clad permittivity) much higher AED is can be achieved for ADBM than both SiNWs and cylindrical air holes. Fig. 4(c) shows the 2D map of the AED as a function of $\xi$ and ridge width, $w_1$. Highest AED which can be achieved for our proposed unoptimized structure using the parameter values given in Fig. 1 is ~16 degree. This is much higher than the AED which can be achieved by both natural materials [4] and all dielectric metamaterials [8]. Fig. 4(d) shows the propagation constant, q as a function of rotation angle, θ for $\xi = .1$ and different values of ridge width. Real valued propagation constants are only found between the limits, $\theta_{min}$ and $\theta_{max}$ which is expected. Also the propagation constant, q has very small variance with the rotation angle.

## CONCLUSION

In conclusion, we have proposed and numerically analyzed an all dielectric biaxial metamateiral which can support lossless DSWs with higher angular existence domain than both natural materials and uniaxial all dielectric metamaterials. Such a metamaterial can be easily constructed by the current fabrication technology [36]. Lossless DSWs supported at the interface of the proposed ADBM and isotropic clad medium can find substantial applications especially in optical sensing, optical interconnects, optical wave guiding etc. It might be superior in performance than traditional plasmonic devices for some applications.

# All Dielectric Biaxial Metamaterial: A New Route to Lossless Dyakonov Waves with Higher Angular Existence Domain

## Supplementary Information

Ayed Al Sayem

Dept. of EEE, Bangladesh University of Engineering and Technology, Dhaka, Bangladesh

## I. OPTICAL PARAMETER RETRIEVAL PROCEDURE

Here, for parameter retrieval procedure, x'=x, y'=y and z'=z. The normal component of the wave vector, $k_x$ and the generalized impedance, $\zeta$ can be calculated by the following equations [S1-S3],

$$k_x d = \pm \cos^{-1}\left(\frac{k_s(1-R^2)+k_c\left(\frac{T}{A}\right)^2}{\left(\frac{T}{A}\right)[ks(1-R)+kc(1+R)]}\right)+2m\pi \tag{S1}$$

with $m \in \mathbb{Z}$.

$$\zeta \pm = \sqrt{\frac{k_s(R-1^2)+k_c\left(\frac{T}{A}\right)^2}{(R+1)^2-\left(\frac{T}{A}\right)^2}} \tag{S2}$$

The sign of eqn. S1 is chosen in such a way that the imaginary part of the normal component of the wave vector, $k_x$ is always positive to ensure exponential decay for wave propagation in the +ve x direction. The value of m is chosen in such a way that it ensures the continuity of both real and imaginary part of the normal component of the wave vector.

For TM (p) polarisation and normal incidence ($k_y = 0$), the effective parallel permittivity in the y direction can be given by,

$$\varepsilon_y = \frac{k_x}{\zeta} \tag{S3}$$

The effective parallel permeability in the y direction can be calculated as,

$$\mu_y = \frac{k_x^2}{\varepsilon_y k_0^2} \tag{S4}$$

Similarly for TE polarisation and normal incidence ($k_z = 0$), the effective parallel permeability in the z direction can be given by,

$$\mu_z = \frac{k_x}{\zeta} \tag{S5}$$

The effective parallel permittivity in the z direction can be calculated as,

$$\varepsilon_z = \frac{k_x^2}{\mu_z k_0^2} \tag{S6}$$

For TM polarisation and at any incidence other than normal incidence ($k_y \neq 0$), the effective perpendicular permittivity in the x direction can be given by,

$$\varepsilon_x = \frac{k_y^2}{\mu_z k_0^2 - \frac{k_x^2}{\varepsilon_y}} \tag{S7}$$

Here we note that though we have conisdered the effective parallel permeabilities throughout our parameter retrieval calculations, we have found that they are near unity. This is expected as all the materials considered in this work are non-magnetic.

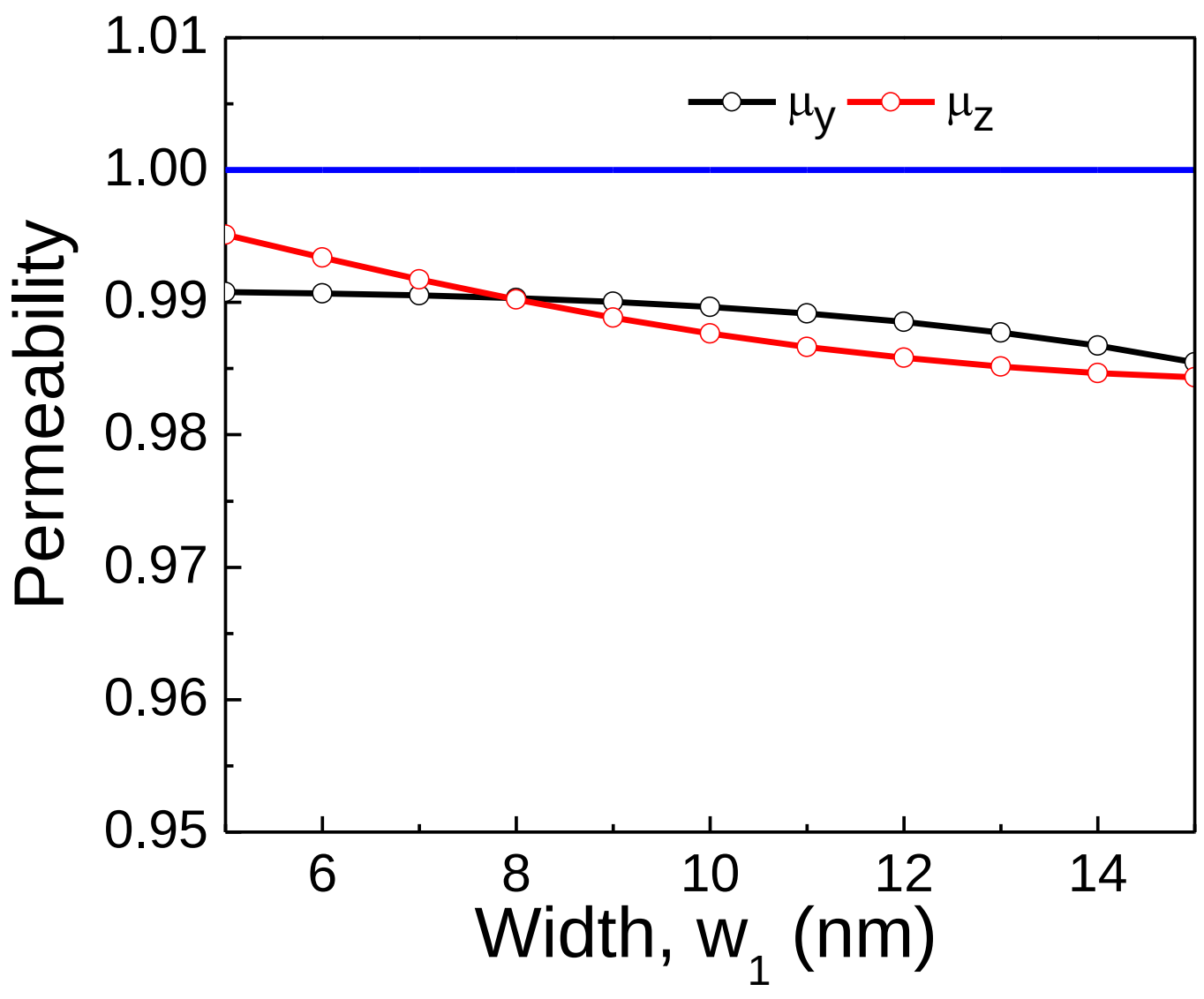

Fig. S1. Retrieved effective parallel permeability $\mu_y$ and $\mu_z$ in the y and z direction respectively as a function of ridge width, $w_1$. All other parameters are same as used in Fig. 1 of the main article. The solid blue line represents the value 1.

Fig. S1 shows the effective parallel permeability values ($\mu_y, \mu_z$) of the proposed ADBM as a function of ridge width. The solid blue line represents the value 1 for comparison. It can be clearly observed from Fig. S1 that the proposed ADBM is almost non-magnetic. All the parallel components of the permeability tensor are very close to unity and slightly diamagnetic [S1].

## II. DSWS AT THE INTERFACE OF UNIAXIAL NANOWIRE METAMATERIAL

Silicon Nanowires in $SiO_2$ host or cylindrical air holes in either Silicon or $SiO_2$ can be treated as an effective uniaxial medium in the long wavelength limit [S4-S6]. The parallel and perpendicular permittivity of such uniaxial medium can be given by [S5],

$$\varepsilon_{||} = f\varepsilon_1 + (1-f)\varepsilon_2 \tag{S8a}$$

$$\varepsilon_{\perp} = \varepsilon_2 \frac{(1-f)\varepsilon_1 + (1+f)\varepsilon_2}{(1-f)\varepsilon_2 + (1+f)\varepsilon_1} \tag{S8b}$$

Where $\varepsilon_1 = \varepsilon_{si}$ and $\varepsilon_2 = \varepsilon_{SiO_2}$ for Silicon Nanowires in $SiO_2$ host and $\varepsilon_1 = 1$ and $\varepsilon_2 = \varepsilon_{Si}$ for cylindrical air holes in Silicon host. We have not considered the case of Silicon nanowires in air. As the DSWs can only propagate along the parallel (making an angle, θ) direction of the nanowires, in such case Silicon nanowires have to be aligned horizontally in air which is difficult from both fabrication and mechanical point of view.

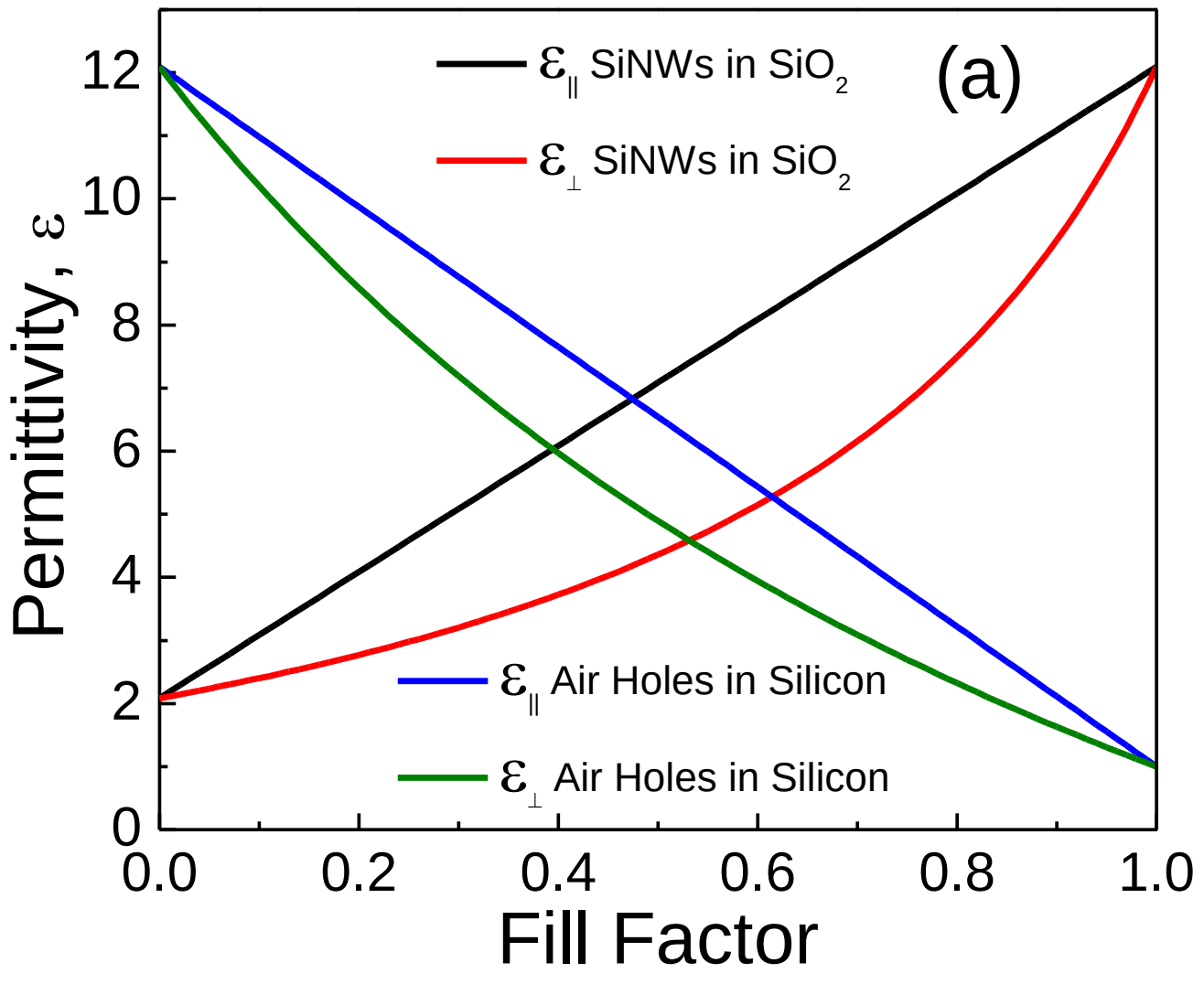

.

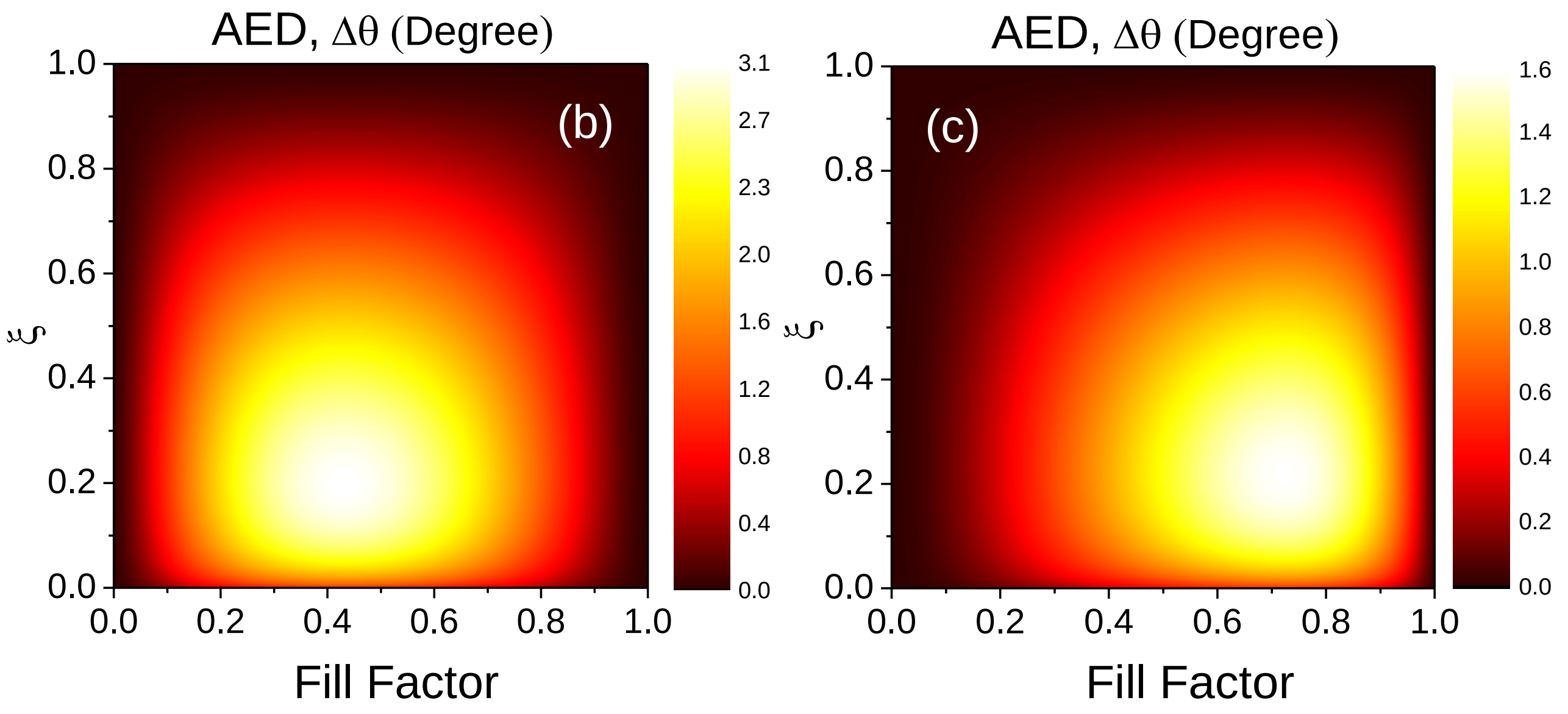

Fig. S2. (a) Parallel and perpendicular permittivity of the SiNWs in $SiO_2$ host and cylindrical air holes in Silicon host (b) 2D map of the AED as a function of fill factor and the parameter, $\xi$ for SiNWs in $SiO_2$ host and cylindrical air holes in Silicon host.

Fig. S2(a) shows the parallel and perpendicular permittivity of the SiNWs in $SiO_2$ host and cylindrical air holes in Silicon host as a function of fill factor. The effective permittivity values vary between the values of Silicon and $SiO_2$ for SiNW metamaterial and Silicon and air for cylindrical air holes metamaterial [S4, S6]. Figs. S2(b) and S2(c) show the 2D map of the AED for SiNWs in $SiO_2$ host and cylindrical air holes in Silicon host as a function of fill factor and the parameter, $\xi$. For SiNWs, the highest value of AED is $3.1^0$ and is achieved at fill factor 0.43 which has been used in Fig. 4(b) of the main article. For cylindrical air holes in Silicon, the highest value of AED is $1.6^0$ and is achieved at fill factor 0.73 which has been used in Fig. 4(b) of the main article.